\pdfoutput=1
\documentclass[]{spie}  

\usepackage{amsmath,amsfonts,amssymb}
\usepackage{graphicx}
\usepackage[colorlinks=true, allcolors=blue]{hyperref}
\usepackage{siunitx}
\usepackage{booktabs}
\usepackage{color}
\usepackage{tablefootnote}

\newcommand{\bicep}{{\textsc {Bicep}}}
\newcommand{\bicepthree}{{\textsc {Bicep3}}}
\newcommand{\planck}{{\textit {Planck}}}

\newcommand{\keck}{{\textit{Keck Array}}}
\newcommand{\biceparray}{{\textsc{Bicep} Array}}
\newcommand{\polarbear}{{\textsc {Polarbear}}}

\newcommand{\ukcmbrts}{$\mu\mathrm{K}_{\mathrm{\mbox{\tiny\textsc cmb}}}\sqrt{\mathrm{s}}$}
\newcommand{\ukcmb}{$\mu\mathrm{K}_{\mathrm{\mbox{\tiny\textsc cmb}}}$}

\title{BICEP Array: a multi-frequency degree-scale CMB polarimeter}

\author[a]{Howard Hui}
\affil[a]{Department of Physics, California Institute of Technology, Pasadena, California 91125, USA}
\author[b]{P.~A.~R.~Ade}
\affil[b]{School of Physics and Astronomy, Cardiff University, Cardiff, CF24 3AA, United Kingdom}
\author[c]{Z.~Ahmed}
\affil[c]{Kavli Institute for Particle Astrophysics and Cosmology, SLAC National Accelerator Laboratory, 2575 Sand Hill Rd, Menlo Park, California 94025, USA}
\author[a]{R.~W.~Aikin}
\author[d]{K.~D.~Alexander}
\affil[d]{Harvard-Smithsonian Center for Astrophysics, Cambridge, Massachusetts 02138, USA}
\author[d]{D.~Barkats}
\author[e]{S.~J.~Benton}
\affil[e]{Department of Physics, University of Toronto, Toronto, Ontario, M5S 1A7, Canada}
\author[f]{C.~A.~Bischoff}
\affil[f]{Department of Physics, University of Cincinnati, Cincinnati, Ohio 45221, USA}
\author[a,g]{J.~J.~Bock}
\affil[g]{Jet Propulsion Laboratory, Pasadena, California 91109, USA}
\author[d]{R.~Bowens-Rubin}
\author[a]{J.~A.~Brevik}
\author[d]{I.~Buder}
\author[h]{E.~Bullock}
\affil[h]{Minnesota Institute for Astrophysics, University of Minnesota, Minneapolis, 55455, USA}
\author[d,i]{V.~Buza}
\affil[i]{Department of Physics, Harvard University, Cambridge, MA 02138, USA}
\author[d]{J.~Connors}
\author[d]{J.~Cornelison}
\author[g]{B.~P.~Crill}
\author[j]{M.~Crumrine}
\affil[j]{School of Physics and Astronomy, University of Minnesota, Minneapolis, 55455, USA}
\author[d]{M.~Dierickx}
\author[w]{L.~Duband}
\affil[w]{Service des Basses Temp\'{e}ratures, Commissariat \`{a} lEnergie Atomique, 38054 Grenoble, France}
\author[i]{C.~Dvorkin}
\author[l,m]{J.~P.~Filippini}
\affil[l]{Department of Physics, University of Illinois at Urbana-Champaign, Urbana, Illinois 61801}
\affil[m]{Department of Astronomy, University of Illinois at Urbana-Champaign, Urbana, Illinois 61801, USA}
\author[j]{S.~Fliescher}
\author[k]{J.~Grayson}
\affil[k]{Department of Physics, Stanford University, Stanford, California 94305, USA}
\author[j]{G.~Hall}
\author[n]{M.~Halpern}
\affil[n]{Department of Physics and Astronomy, University of British Columbia, Vancouver, British Columbia, V6T 1Z1, Canada}
\author[d]{S.~Harrison}
\author[a,g]{S.~R.~Hildebrandt}
\author[o]{G.~C.~Hilton}
\affil[o]{National Institute of Standards and Technology, Boulder, Colorado 80305, USA}
\author[k,c,o]{K.~D.~Irwin}
\author[k]{J.~Kang}
\author[d,p]{K.~S.~Karkare}
\affil[p]{Kavli Institute for Cosmological Physics, University of Chicago, Chicago, IL 60637, USA}
\author[k]{E.~Karpel}
\author[q]{J.~P.~Kaufman}
\affil[q]{Department of Physics, University of California at San Diego, La Jolla, California 92093, USA}
\author[q]{B.~G.~Keating}
\author[a]{S.~Kefeli}
\author[k]{S.~A.~Kernasovskiy}
\author[d,i]{J.~M.~Kovac}
\author[k,c]{C.~L.~Kuo}
\author[j]{K.~Lau}
\author[r]{N.~A.~Larsen}
\affil[r]{Kavli Institute for Cosmological Physics, University of Chicago, Chicago, IL 60637, USA}
\author[r]{E.~M.~Leitch}
\author[a]{M.~Lueker}
\author[g]{K.~G.~Megerian}
\author[a]{L.~Moncelsi}
\author[s]{T.~Namikawa}
\affil[s]{Leung Center for Cosmology and Particle Astrophysics, National Taiwan University, Taipei 10617, Taiwan}
\author[e,t]{C.~B.~Netterfield}
\affil[t]{Canadian Institute for Advanced Research, Toronto, Ontario, M5G 1Z8, Canada}
\author[g]{H.~T.~Nguyen}
\author[a,g]{R.~O'Brient}
\author[k,c]{R.~W.~Ogburn~IV}
\author[f]{S.~Palladino}
\author[h,j]{C.~Pryke}
\author[d]{B.~Racine}
\author[d]{S.~Richter}
\author[h]{R.~Schwarz}
\author[a]{A.~Schillaci}
\author[u,p]{C.~D.~Sheehy}
\affil[u]{Physics Department, Brookhaven National Laboratory, Upton, NY 11973}
\author[a]{A.~Soliman}
\author[d]{T.~St.~Germaine}
\author[a,g]{Z.~K.~Staniszewski}
\author[a]{B.~Steinbach}
\author[b]{R.~V.~Sudiwala}
\author[a,q]{G.~P.~Teply}
\author[k,c]{K.~L.~Thompson}
\author[k]{J.~E.~Tolan}
\author[b]{C.~Tucker}
\author[g]{A.~D.~Turner}
\author[f]{C.~Umilt\`{a}}
\author[v,p]{A.~G.~Vieregg}
\affil[v]{Department of Physics, Enrico Fermi Institute, University of Chicago, Chicago, IL 60637}
\author[a]{A.~Wandui}
\author[g]{A.~C.~Weber}
\author[b]{D.~V.~Wiebe}
\author[h]{J.~Willmert}
\author[d,i]{C.~L.~Wong}
\author[k,c,r]{W.~L.~K.~Wu}
\author[c]{E.~Yang}
\author[c,k]{K.~W.~Yoon}
\author[a]{C.~Zhang}

\authorinfo{Further author information send correspondence to H. Hui\\H. Hui: E-mail: hhui@caltech.edu, Telephone: 1 626 395 2023}

\begin{document} 
\maketitle

\begin{abstract}
\biceparray\ is the newest multi-frequency instrument in the \bicep/\keck\ program.
It is comprised of four 550\,mm aperture refractive telescopes observing the polarization of the cosmic microwave background (CMB) at 30/40, 95, 150 and 220/270\,GHz with over 30,000 detectors.
We present an overview of the receiver, detailing the optics, thermal, mechanical, and magnetic shielding design.
\biceparray\ follows \bicepthree\,'s modular focal plane concept, and upgrades to 6" wafer to reduce fabrication with higher detector count per module.
The first receiver at 30/40\,GHz is expected to start observing at the South Pole during the 2019-20 season.
By the end of the planned \biceparray\ program, we project $ \sigma(r) \sim 0.003$, assuming current modeling of polarized Galactic foreground and depending on the level of delensing that can be achieved with higher resolution maps from the South Pole Telescope.

\end{abstract}


\section{Introduction}

Measurements of the polarization of the Cosmic Microwave Background provide key information to further our understanding of the early universe.
The $\Lambda$CDM model predicts an $E$-mode polarization pattern in the CMB at the level of a few $\mu$K as well as an arc-minute $B$-mode polarization arising from gravitational lensing of $E$-modes by the large-scale structure of the universe.
Inflationary gravitational waves may be a source of degree-scale $B$-mode polarization and a detection of such signal can be used to constrain the tensor-scalar ratio $r$ and place limits on the energy scale and potential of Inflation~\cite{kamionkowski2016quest}.
While classes of Inflation models could generate undetectably low levels of gravitational waves, a detection of $B$-mode polarization generated by primordial gravitational waves would be direct evidence for the theory of Inflation.
However, in order to disentangle a potential CMB signal from polarized Galactic dust and synchrotron foregrounds, we need to probe the polarization of the CMB at multiple frequencies with high sensitivity.

\begin{figure}[h]
\centering
\includegraphics[width=0.75\textwidth]{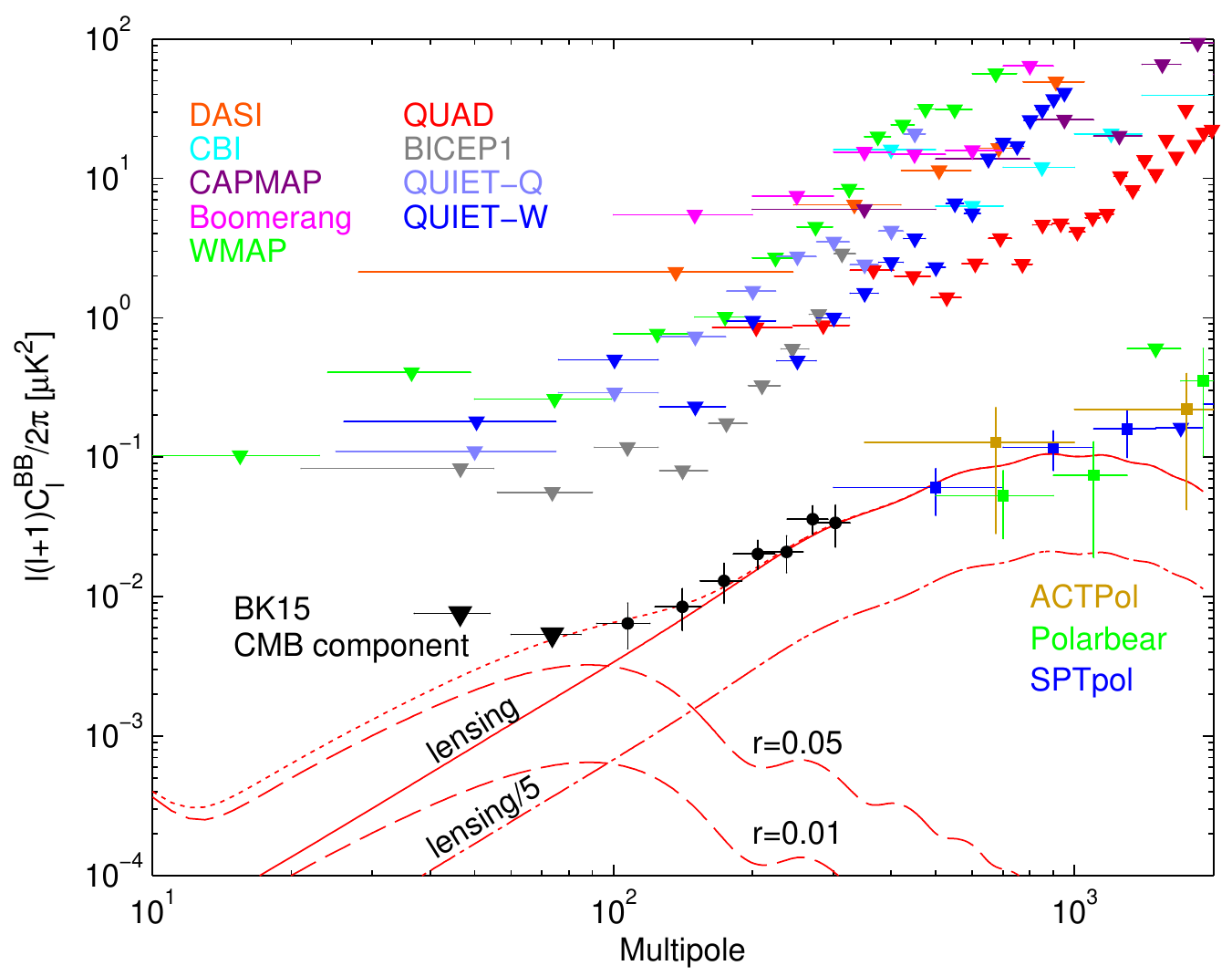}
\caption[Published $B$-mode polarization measurements]{
Published $B$-mode polarization measurement by different experiments.
$B$-mode polarization from gravitational lensing has been detected by the \bicep/\keck, SPT, \polarbear, and ACT collaborations.
}
\label{fig:bk_15_status}
\index{figure}
\end{figure}

\begin{figure}
   \begin{center}
   \includegraphics[width=0.75\textwidth]{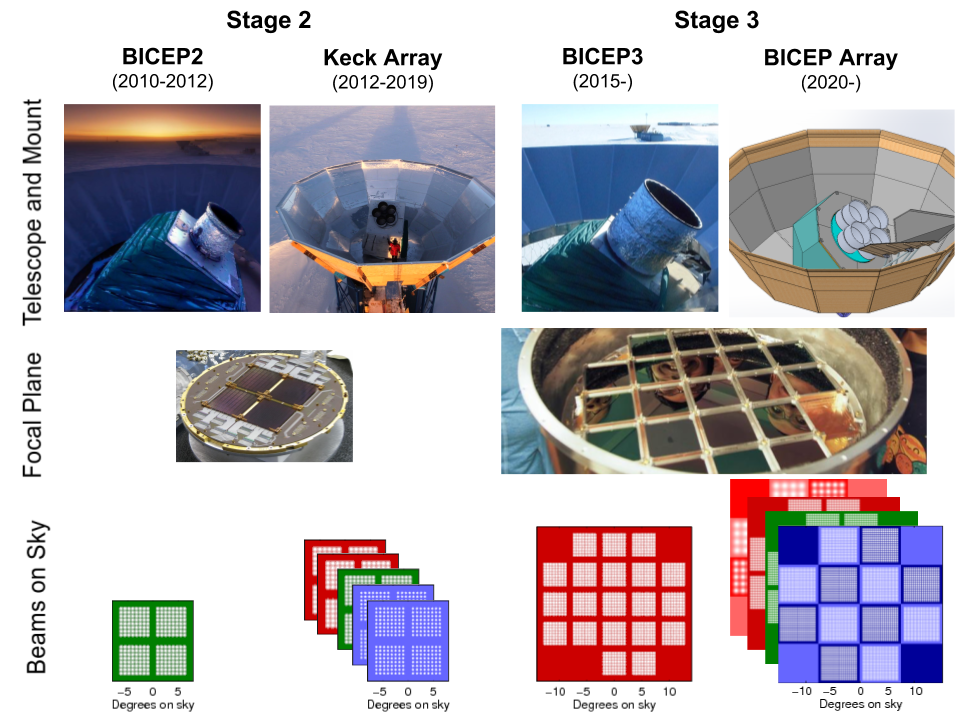}
   \end{center}
   \caption{The progression of the \bicep/\keck\ program leading to the \biceparray.
   Bottom row: the beam patterns of the focal planes on the sky shown on a common scale.
   Each square represents a single receiver, and the colors indicate different observing frequencies in pink (30/40\,GHz), red (95\,GHz), green (150\,GHz), and blue (220/270\,GHz).}
   \label{fig:bk_program} 
   \end{figure}

The current constraint on tensor-to-scalar ratio is $r_{0.05}<0.06$ at 95\% confidence from \bicep/\keck\ data in conjunction with \planck\ temperature measurements~\cite{BK15} (Figure~\ref{fig:bk_15_status}).
Over the past 10 years, our experimental strategy of utilizing small-aperture, cold, refracting telescopes has proven to be successful to probe the degree-scale polarization of the CMB.
\bicep2\ observed the sky with 500 antenna-coupled transition-edge sensor (TES) bolometers at 150\,GHz from 2010 to 2012, and reported a 5\,$\sigma$ excess of $B$-mode power over the base lensed-$\Lambda$CDM model in the range $30<l<150$~\cite{ade2014detection}.
The \keck\ consists of five 25\,cm aperture receivers, each similar to \bicep2, started observations at 150\,GHz in 2012.
A joint analysis with \planck\ indicated the signal reported from \bicep2 is consistent with polarized emission from Galactic dust~\cite{BKP}.
The interchangeable \keck\ receivers allowed us to diversify the frequency coverage by rapidly switching each receiver to 95, 220, and 270\,GHz.
The latest instrument in our program, \bicepthree, replaced \bicep2 on its mount in 2015.
\bicepthree\ uses a 0.52\,m telescope and 2500 detectors operating at 95\,GHz to realize an on-sky instantaneous sensitivity of 6.7\,\ukcmbrts \cite{kang2018bicep3}.
Figure~\ref{fig:bk_program} shows the progression of the \bicep/\keck\ program, to larger apertures, larger focal planes, and wider frequency coverage.

\begin{figure}
  \centering
  \includegraphics[width=1\textwidth]{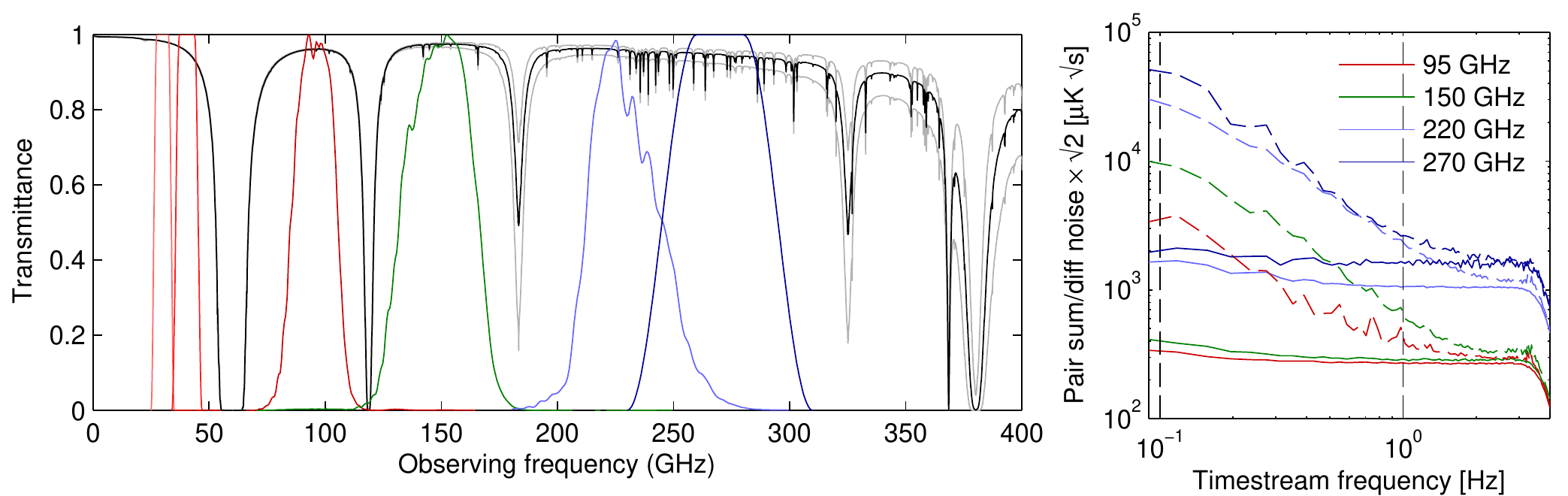}
  \caption{
  {\it Left:} Comparison of atmospheric transmission at the South Pole  with the bandpasses of \bicep/\keck\ and \biceparray.
  Median atmospheric transmission during the observing season is shown in black, bracketed by the 10\textsuperscript{th} and 90\textsuperscript{th} percentiles.
  Transmission drops only slightly across 200--300\,GHz, making dust observations in the upper part of this window effective, with similar dust sensitivity to the 220 GHz band.
{\it Right:} Minimally processed timestream pair-sum and pair-difference noise spectra from \keck.
The stable Antarctic atmosphere enables observations at all of these frequencies that are low-noise across the indicated science band from 0.1--1\,Hz, corresponding to $25\lesssim\ell\lesssim250$.
}
  \label{fig:atmos}
  \index{figure}
\end{figure}

\biceparray\ adopts the same interchangeable concept used in \keck\ and is comprised of four \bicepthree-class receivers, each optimized for a atmospheric window in the frequency range from 30 to 270\,GHz (Figure~\ref{fig:atmos}).
The highest and lowest frequency receivers incorporate two bands within an atmospheric window, operating at 30/40~GHz and 220/270~GHz, by shifting bandpass in alternating focal plane modules over the focal plane (Figure~\ref{fig:bk_program}).
The splitting of the band provides more information on polarized Galactic synchrotron and dust emission to test the parameters of the foreground model.

The \keck\ telescope mount will be replaced by the new \biceparray\ mount\cite{crumrine2018biceparray} at the end of 2019.
\biceparray\ receivers will be installed in the new mount in a staged approach over the next few years, with the first 30/40\,GHz receiver to be deployed at the end of 2019, followed by the 150, 95 and 220/270\,GHz receivers.
We are currently planning on continuing observations with a subset of the \keck\ receivers installed in open slots until they are filled by available \biceparray\ receivers.
The parameters of the \keck, \bicepthree\ and \biceparray\ receivers are given in Table~\ref{tab:rxs}.

\begin{table}
\small
\begin{center}
\begin{tabular}[c]{|l|c|c|c|c|}
\hline
Receiver       & Nominal  & Nominal Single       & Beam       & Survey Weight \\
Observing Band & Number of& Detector NET         & FWHM       & Per Year      \\
(GHz)          & Detectors& (\ukcmbrts)          & (arcmin)   & (\ukcmb)$^{-2}$ yr$^{-1}$\\
\hline\hline   						
\keck    &          &                      &            &               \\
\ \ \ 95           & \textbf{288} & 288                  & \textbf{43}    & \textbf{24,000}   \\
\ \ \ 150          & \textbf{512} & 313                  & \textbf{30}    & \textbf{30,000}   \\
\ \ \ 220          & \textbf{512} & 837                  & \textbf{21}    & \textbf{2,000}    \\
\ \ \ 270          & \textbf{512}      & 1310                 & \textbf{17}         & 800           \\
\hline\hline   						
\bicepthree    &          &                      &            &               \\
\ \ \ 95           & \textbf{2560}& 288                  & \textbf{24}    & \textbf{213,000}       \\
\hline\hline   						
\biceparray     &          &                      &            &               \\
$\big \langle \hspace{-3pt} \begin{array}{l} 30 \\ 40 \end{array}$   & $\begin{array}{l} 192 \\ 300 \end{array}$
  & $\begin{array}{l} 221 \\ 301 \end{array}$ & $\begin{array}{l} 76 \\ 57 \end{array}$
  & $\begin{array}{l} 19,500 \\ 20,500 \end{array}$       \\
\ \ \ $95$           & $3456$     & $288$                  & $24$         & $287,000$       \\
\ \ \ $150$          & $7776$     & $313$                  & $15$         & $453,000$       \\
$\big \langle \hspace{-3pt} \begin{array}{c} 220 \\ 270 \end{array}$   & $\begin{array}{c} 8112 \\ 13068 \end{array}$
  & $\begin{array}{c} 837 \\ 1310 \end{array}$ & $\begin{array}{c} 11 \\ 9 \end{array}$
  & $\begin{array}{c} 37,000 \\ 15,000 \end{array}$       \\
\hline\hline
\end{tabular}
\end{center}
\caption[Receiver parameters and sensitivity for \bicep\ program]{Receiver parameters as used in sensitivity projections.
Boldface numbers are actual/achieved quantities for existing receivers.
The remaining values in the survey weight column are scaled from the achieved survey weights using only the ratio of the number of detectors, plus, if necessary to change frequency, the ratio of nominal NET values squared.
In 2017 the 270\,GHz \keck\ receiver realized single-detector NETs of 1310~\ukcmbrts\ from pair differences in the best channels, but with a large dispersion due to excess detector noise in many detectors.
An improved 270\,GHz focal plane is fielded in 2018.}
\label{tab:rxs}
\end{table}

\section{Bicep Array instrument overview}
\subsection{Receiver overview}

\begin{figure}
  \centering
  \includegraphics[width=1\textwidth]{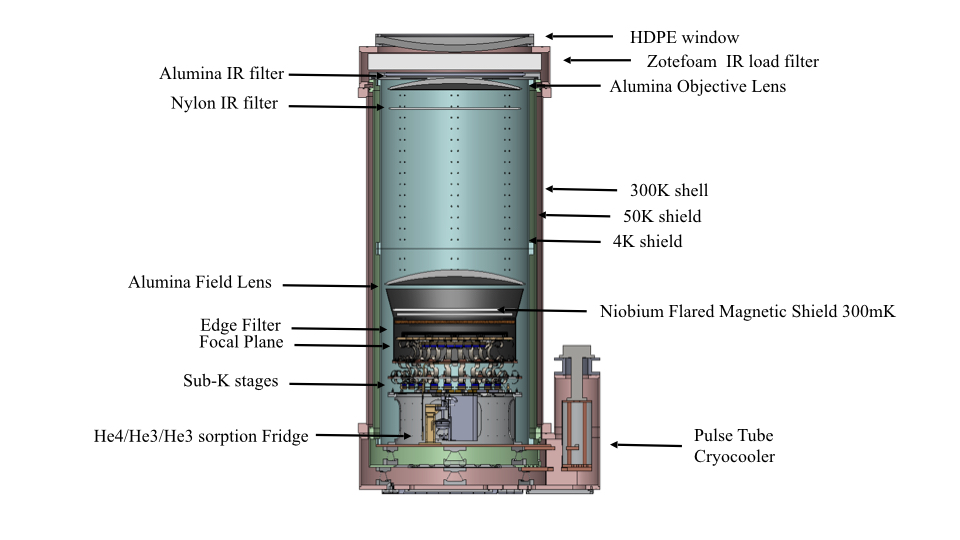}
  \caption[\biceparray\ receiver cutaway]{
Cross-sectional view of the \biceparray\ receiver.
The alumina lenses, nylon filter, and aperture stop are precision mounted in the receiver and cooled to 4~K.
The focal plane assembly houses the detector modules is cooled to 250~mK by the sub-Kelvin sorption refrigerator and surrounded by a superconducting Nb magnetic shield. 
}
  \label{fig:ba_cryostat}
  \index{figure}
\end{figure}

\biceparray\ is largely based on the successful design of \bicepthree \cite{2016JLTPB3}.
Each receiver is housed in a custom-designed vacuum cryostat 2.1\,m tall and 0.9\,m in diameter (excluding the additional envelopes of the multi-channel readout electronics and the pulse-tube cooler).
Figure~\ref{fig:ba_cryostat} shows a cross-section of the receiver.
The vacuum jacket and the 50\,K stages are constructed with a short base stage, and long cylinders for the main section.
This allows us to lift off the outer cylinders from the base plate without removing the cabling, thermal joints and focal plane structure.
The top of the vacuum jacket is capped by a vacuum window (made out of HDPE for the 30/40\,GHz receiver; solutions for the higher frequencies are discussed in a companion paper\cite{dierickx2018window}), and a stack of Zotefoam\textsuperscript{\textregistered} infrared filters behind it.
An IR-absorptive alumina filter\cite{inoue2014cryogenic} is installed at the top of the 50\,K stage.
 
The 4\,K stage is sectioned into two lengthwise segments for ease of access during integration; the top optics section houses the optical elements, including the two alumina lenses, and potentially a nylon filter, to absorb long-wavelength infrared emission before reaching the sub-K focal plane.
The lower camera section houses the sub-kelvin cryogenic system, time-domain multiplexed SQUID readouts, and the sub-kelvin focal plane with a thermo-mechanical truss structure.

The 50\,K volume is supported by G-10 trusses, providing robust structural support while maintaining low thermal conductivity between temperature stages.
The 4\,K and sub-Kelvin structures are supported by carbon fiber trusses because of its high ratio of stiffness to thermal conductivity at low temperature\cite{Runyan2008}.

\subsection{Thermal Architecture}

The radiative loading from the window and warmer stages in \biceparray\ dominates over conducted loading inside the receiver, thus the receiver's thermal architecture is focused on minimizing the instrument internal loading on the detectors.
The 50\,K and 4\,K stages are heat sunk to the first and second stages of a Cryomech\footnote{Cryomech, Inc. (\texttt{www.cryomech.com})} PT-415 cooler, with cooling capacity of 40\,W and 1.5\,W, respectively.

The 56\,cm aperture window lets $>100$\,W of infrared radiation into the receiver.
The current baseline material for the vacuum window is 1"-thick HDPE at 30/40\,GHz, where HDPE has negligible in-band emission, but at higher frequencies thinner windows using polyethylene fiber composites will reduce the in-band optical loading and thereby significantly improve mapping speed\cite{dierickx2018window}.
Within the receiver, we use a combination of Zotefoam\textsuperscript{\textregistered} filters, alumina optics and a nylon filter to reduce infrared loading to an acceptable level for the sub-Kelvin stages to function.
Combining actual measurements from \bicepthree\ with a thermal model, the total loading on the 50\,K and 4\,K stages is estimated to be about 21\,W and 0.38\,W, respectively, implying actual base temperatures of 34\,K and 3.1\,K at these stages.

\begin{figure}
  \centering
  \includegraphics[width=0.5\textwidth]{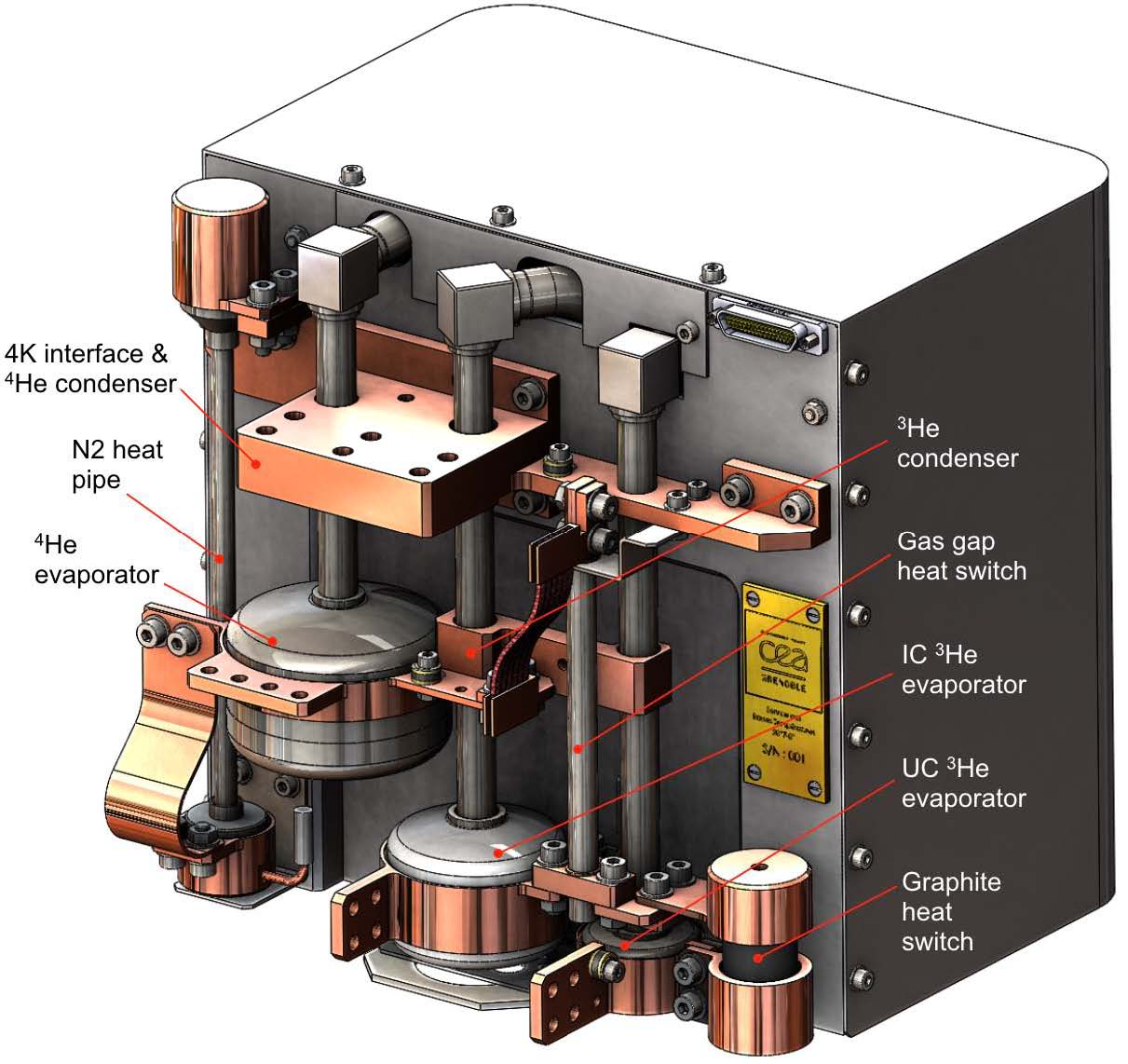}
  \caption[CAD model of the \biceparray\ fridge]{
Computer-aided design model of the \biceparray\ fridge. Each stage is equipped with a removable heat switch connected to the higher-temperature stage to accelerate the initial cool down. (Credits: CEA Grenoble\cite{ICC_Duband})
}
  \label{fig:ba_fridge}
  \index{figure}
\end{figure}

Sub-Kelvin cooling for the detectors is provided by a three-stage helium (4He/3He/3He) sorption fridge from CEA Grenoble\cite{ICC_Duband} (Figure~\ref{fig:ba_fridge}) at 2\,K (4He stage), 340\,mK (intermediate cold), and 250\,mK (ultra-cold).
With estimating 20 and 15~$\mu$W of optical loading at 340 and 230~mK stages, it has sufficient cooling capacity to have a 3-day uninterrupted observing schedule between thermal cycles.
The detector modules and focal plane are heat sunk to the ultra-cold stage via a flexible high-purity copper-foil heat strap.
The strap connects to the focal plane through a stainless steel block. 
The temperature of the strap is actively regulated, and the block serves as a passive low-pass thermal filter to attenuate thermal noise from the control circuit.
The estimated conductive loading on the ultra-cold stage is 0.25\,$\mu$W (not including radiative contributions, see Table~\ref{tab:ba_subK_load}).

\begin{table}
  \center
  \begin{tabular}{l r r r} 
    \multicolumn{4}{c}{Conducted load budget for the focal plane assembly}\\
    \toprule 
    					&	2~K ($\mu$W)	&	340~mK ($\mu$W)	&	230~mK (nW)\\
    \toprule 
    16 NbTi cables		&	110.7		&	27.7				&	99.6	\\
    4 Manganin cables	&	3.2			&	0.9				&	3.5		\\
    Cernox cables		&	11.0			&	2.1				&	3.0		\\
    Heater cables		&	4.6			&	0.8				&	2.3		\\
    Carbon fiber trusses		&	44.0			&	16.9				&	47.1	\\
    Aluminize mylar shield		&	52.9			&	23.4				&	97.8	\\
    \midrule
    Total				&	226.5		&	71.7				&	253.2	\\
    \bottomrule
    \vspace{1mm}
  \end{tabular}
  \caption[Sub-Kelvin loading from conduction for \biceparray.]{Sub-Kelvin loading from conduction for \biceparray.
  This calculation is based on the maximum cable count for the high-frequency receiver. The loading is expected to be smaller with less readout cables for the lower frequency receivers.}
  \label{tab:ba_subK_load}
\end{table}

\subsection{Housekeeping electronics}

\begin{figure}
  \centering
  \includegraphics[width=0.75\textwidth]{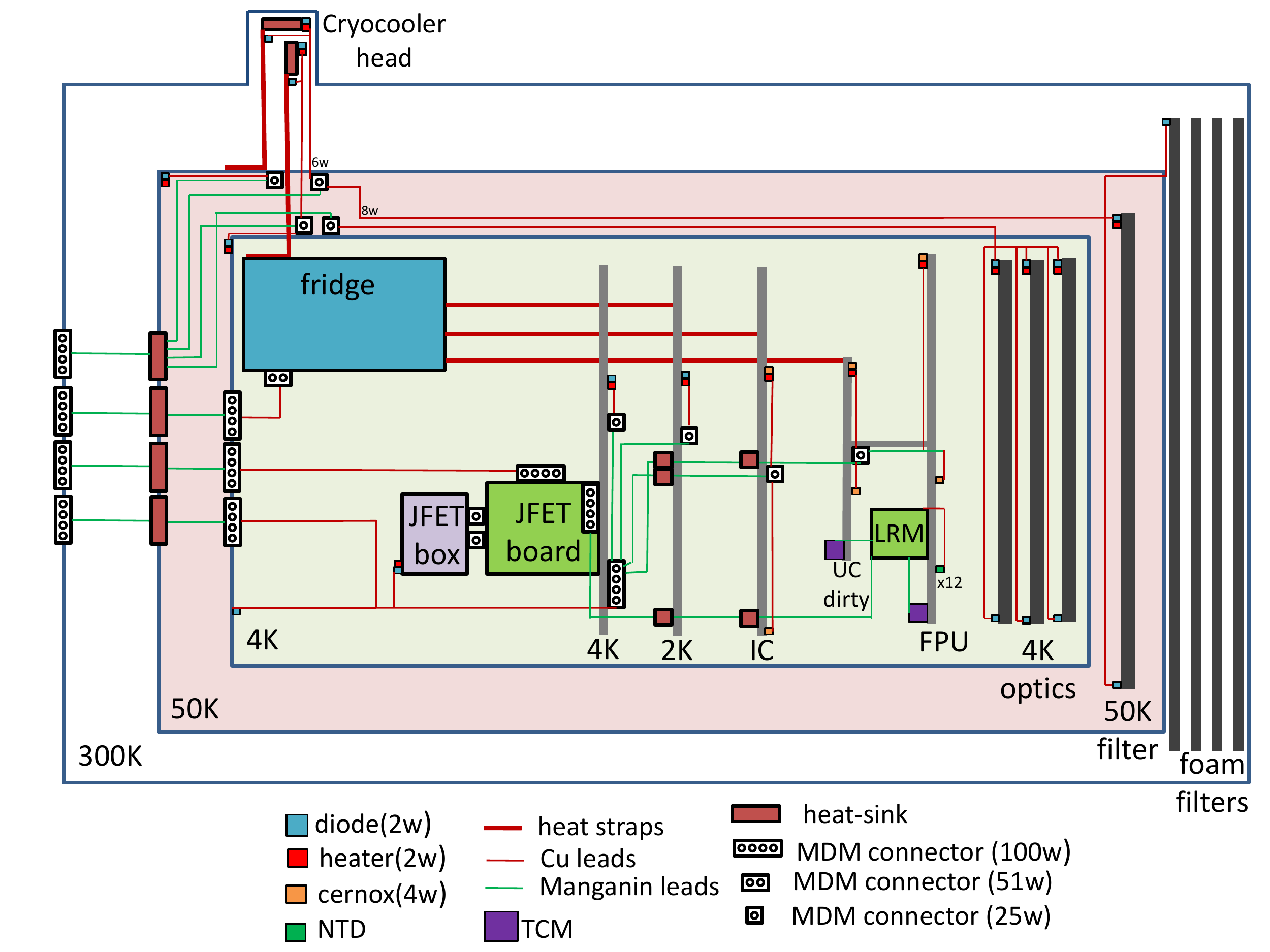}
  \caption[\biceparray\ housekeeping layout]{
\biceparray\ cold housekeeping layout\cite{hui2018measuring}
}
  \label{fig:ba_hk_layout}
  \index{figure}
\end{figure}

Thermal monitoring for the cryostat is done using calibrated diodes and resistance thermometers (cernox\textsuperscript{\textregistered}) at the radiation shields, critical cryogenic junctions, and some of the optical elements.
The focal plane temperature is regulated by passive and active filtering as described above, similar to the approach used on \keck\ and \bicepthree.
Active control is implemented in a feedback loop using Neutron transmutation doped (NTD) Ge thermometers and a resistive heater.
The design uses the same precision NTD Ge thermometry to passively monitor focal plane temperatures.
Figure~\ref{fig:ba_hk_layout} shows the cold housekeeping layout in \biceparray.

Warm housekeeping electronics consist of an analog electronics enclosure, which interfaces to the cryostat and is responsible for biasing thermometers and pre-amplifying their signals, and a commercial electronics crate\footnote{United Electronic Industries (\texttt{www.ueidaq.com})}, which digitizes thermometry signals and sets heater voltages.
The housekeeping electronics support up 64 thermometry channels and 32 heater channels for each \biceparray\ receiver.

\subsection{Optics}
\begin{figure}
  \centering
  \includegraphics[width=0.85\textwidth]{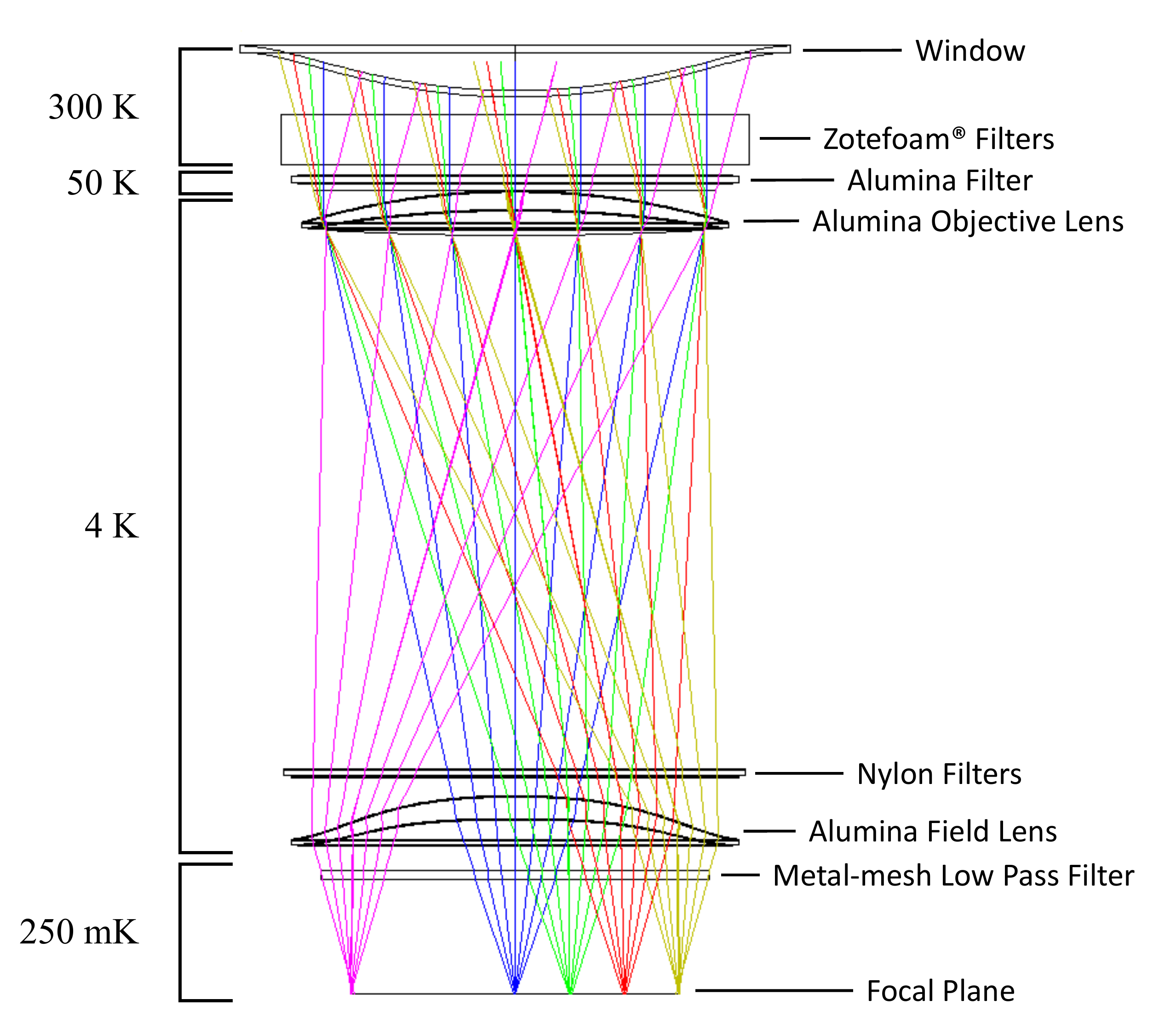}
  \caption[\biceparray\ optical diagram]{
\biceparray\ optical diagram.
All components except the zotefoam filters are anti-reflection coated to provide minimal reflection.
The lenses are cooled to 4~K, and the focal plane is actively temperature controlled to provide low and stable optical loading.
The radially symmetric design allows well-matched beams for two idealized orthogonally polarized detectors at the focal plane.
}
  \label{fig:ba_raytrace}
  \index{figure}
\end{figure}

Following the same concept as in previous telescopes, \biceparray\ uses a simple diffraction-limited refractor, which provides a telecentric and flat focal surface with minimum aberrations over a wide field of view.
The telescope has a mean $f$-ratio of $f$/1.6, and the alumina lenses are 650\,mm in diameter, with a clear aperture of 550\,mm and a field of view of 29.6~degree.
Figure~\ref{fig:ba_raytrace} shows the optical diagram of the 30/40\,GHz design.
The lenses in \biceparray\ are cooled to 4\,K to minimize loading on the detectors.
The 4\,K space between the objective and the field lens is covered in epoxy-encapsulated Eccosorb\textsuperscript{\textregistered} HR-10 microwave absorber to suppress far-sidelobes reflections.
The same Eccosorb\textsuperscript{\textregistered} is also used to define the 550~mm optical stop for the system, located behind the objective lens (in the anti-sky direction).
Simulations show the Gaussian beam width is $\sigma\sim76'/57'$ at 30/40\,GHz. 
Given the similarity to the \bicepthree\ design, we expect the actual values to compare well to the design model.

\subsection{Focal plane and detector module}

The sub-Kelvin structure is located on top of the 4\,K base plate, above the sorption fridge. It is separated into three thermal, ``wedding-cake'' stages at 2\,K, 350\,mK, and 260\,mK.
Each stage provides radiative shielding and room for cable heat-sinking to the respective fridge stages, allowing low loading environment for the focal plane and detectors.
The copper focal-plane plate is mounted at the 250~mK stage, separated by a low-pass stainless steel filter and carbon fiber trusses.

Extensive finite-element simulations were performed on the truss structure to estimate its maximum static load as well as its vibrational resonances.
The minimum resonance frequency for the carbon fiber truss structure is 55\,Hz.
The maximum displacement of the truss structure when subject to a 90\,degree gravity load is less than 70\,$\mu$m at the focal plane.
The first buckling mode of the structure, with a 45~degrees gravity load, is found to be more than 9~times the nominal load in correspondence of the carbon fibers truss elements.
In practice the glued joints are usually weaker than in simulation, so all truss elements will go through warm and cold stress tests before installation.

\begin{figure}
  \centering
  \includegraphics[width=0.5\textwidth]{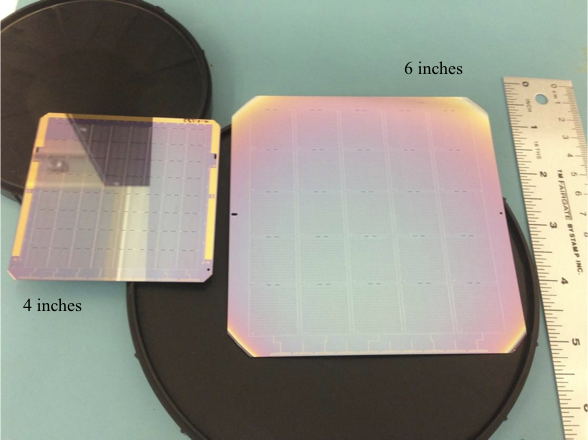}
  \caption[3" vs 6" detector wafer]{
4" and 6" detector wafers used in \bicepthree\ and \biceparray, respectively.
The larger wafers reduce fabrication with higher detector count per module.
Larger optical throughout 
The 95\,GHz \biceparray\ receiver will house $\sim$3500 detectors compared to 2560 detectors in \bicepthree.
}
  \label{fig:ba_wafer}
  \index{figure}
\end{figure}

\begin{figure}
  \centering
  \includegraphics[width=1\textwidth]{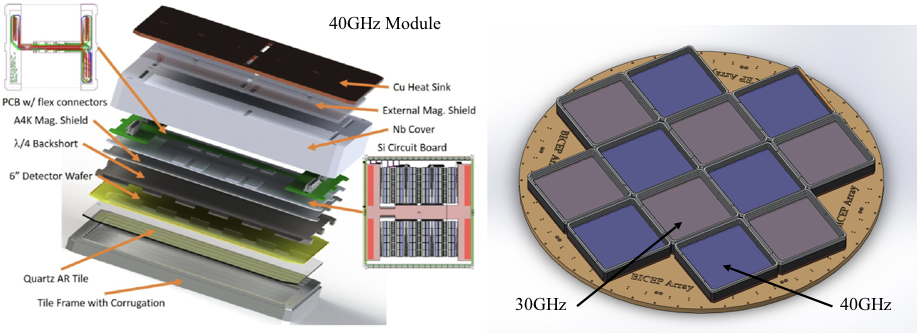}
  \caption[\biceparray\ module and focal plane]{
{\it Left}: An exploded view of the \biceparray\ module.
The design is based on \bicepthree\ but scaled to house a larger detector wafer.
{\it Right}: The 30/40\,GHz focal plane layout.
The receiver has six 30\,GHz and six 40\,GHz modules laid out in a checker-board pattern.

}
  \label{fig:ba_module}
  \index{figure}
\end{figure}

Twelve detector modules are tiled onto the focal plane, each containing 32 to 2178 antenna-coupled transition-edge sensors\cite{Roger2015}, depending on the observing frequency (Table~\ref{tab:rxs}).
The module is based on the successful \bicepthree\ design\cite{hui2016bicep3}, except the detectors are now fabricated on 6" silicon wafers, instead of the previous 4" wafers to expedite lithographic fabrication with higher detector count per module (Figure~\ref{fig:ba_wafer}).
The total number of module is also reduced in the focal plane, simplifies the assembly procedure and reduces the non-optically illuminated metal frame area between modules. 
The first-stage SQUID readout multiplexing chips are housed in a silicon/aluminum-nitride circuit board inside the module.
The housing is constructed with superconducting niobium and aluminum.
Combining with the high-$\mu$ A4K sheet inside the module, it is designed to achieve high magnetic shielding performance (Section~\ref{magnetic shield}).
Figure~\ref{fig:ba_module} shows the exploded view of the module design.

Interaction between the edge antennas and the metal frame around the module cause differential ellipticity, resulting in potential temperature to polarization leakage in the CMB maps. We designed corrugated walls to minimize this effect. The 30/40\,GHz module implements a novel double-corrugation wall that mitigates the differential ellipticity caused by the metal frame over 57\% bandwidth, from 25\,GHz to 45\,GHz~\cite{ahmed2018corr}.

\subsection{Detector Readout}

\begin{figure}
  \centering
  \includegraphics[width=1\textwidth]{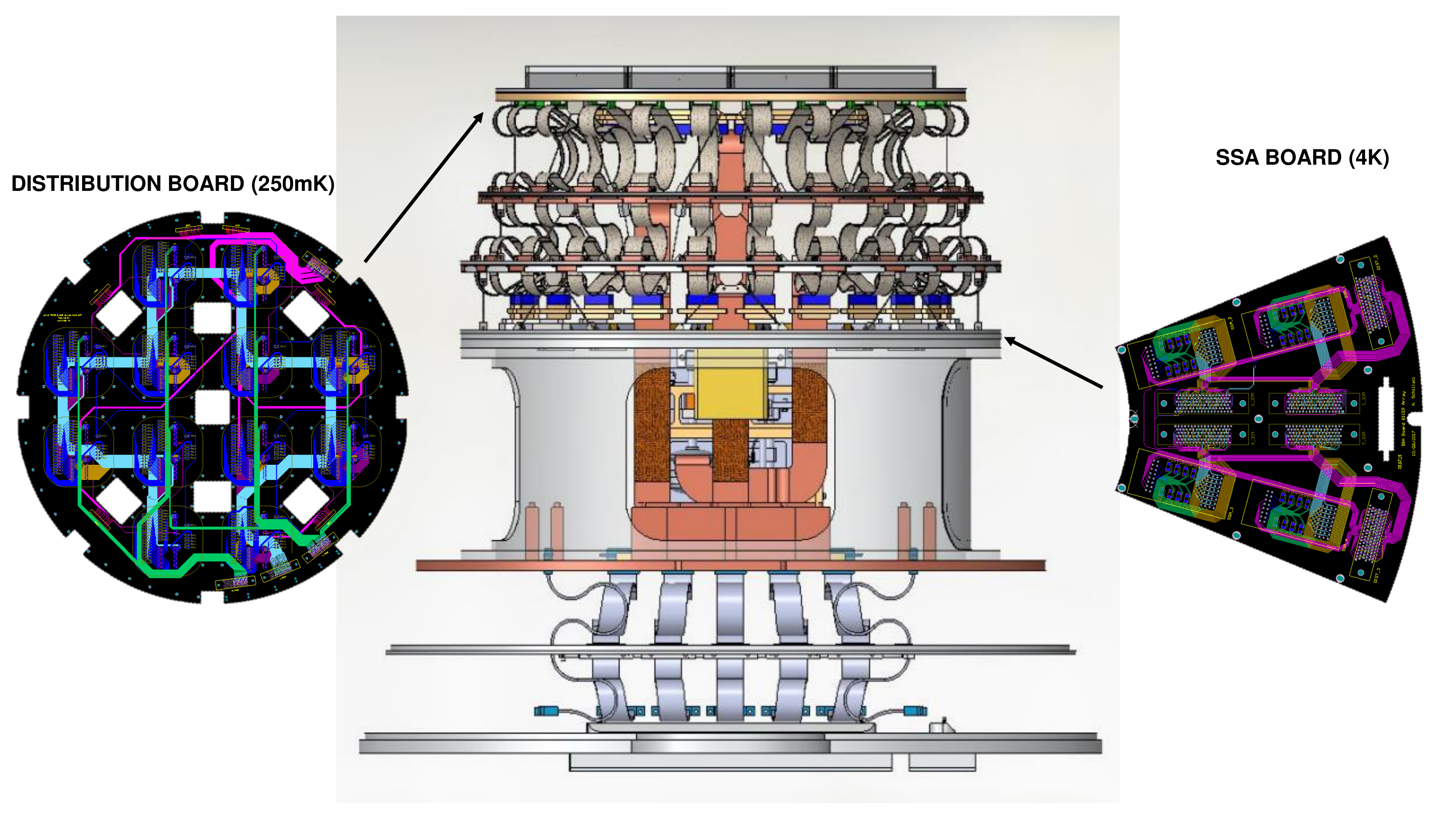}
  \caption{The distribution board gathers all the detector readout cabling and distributes them to the corresponding detector module at 250\,mK.
  The SSA circuit board is located at 4\,K, containing the series SQUID arrays.
  Each MCE connects to a single SSA board via five 100-way cables.
  The 30/40\,GHz receiver contains one SSA circuit board, with one MCE; the 95\,GHz receiver contains three identical boards and the 150\,GHz receiver will have six.
}
  \label{fig:ba_readout_board}
  \index{figure}
\end{figure}

\biceparray\ uses a time-domain multiplexing (TDM) readout system identical to \bicepthree\ for the 30/40, 95, and 150\,GHz receivers \cite{hui2016bicep3}.
Table~\ref{tab:ba_tmux} shows the multiplexing factors and arrangement for each receiver.
The first stage SQUIDs are integrated into the detector module.
A circuit board (distribution board) is mounted underneath the copper focal plane, gathering all the readout cabling and distributing to the corresponding detector module.
The housekeeping electronics and series SQUID array (SSA) amplification stage are placed on the 4\,K base plate circuit boards (Figure~\ref{fig:ba_readout_board}).
The amplified detector signals are digitized and read out using the warm Multi-Channel Electronics (MCE\cite{Battistelli2008}), which also provide SQUIDs and detector bias, and apply feedback for control.
The large number of MCEs needed at 150\,GHz required a relatively minor mechanical redesign of the MCE boxes in order to fit within the space constraints imposed by the telescope mount.

The 220/270\,GHz receiver, with more than 22,000 detector channels, will require a new readout technology.
Thermal kinetic inductance detectors~\cite{obrient2018tkid} (TKID), and microwave SQUID multiplexing~\cite{henderson2018umux} ($\mu$MUX) technologies provide the higher multiplexing factors to handle such large formats and are being developed for the high frequency receiver within the collaboration.

\begin{table}
  \center
  \begin{tabular}{l c c c} 
    \toprule 
    Frequency			&	30/40~GHz 	&	95~GHz	&	150~GHz			\\
    \midrule 
    \# Detector Tiles			&	12			&	12		&	12				\\
    \# Detectors		&	192 + 300	&	3456		&	7776				\\
	\# Detectors/Tile			&	32 + 50		&	288		&	648				\\
	\# SQUID MUX11 chips/Tile 	&	6			&	28		&	64				\\
	\# MCE 				&	1			&	3		&	6				\\
	\# Columns/MCE 		&	24			&	28		&	32				\\    
	\# Rows (multiplexing factor) 			&	33			&	43		&	42				\\    
    \bottomrule
    \vspace{1mm}
  \end{tabular}
  \caption{Multiplexing scheme for the 30/40, 95, and 150~GHz receivers in \biceparray\.
  The fourth 220/270 GHz receiver will use RF multiplexing with higher multiplexing factors.}
  \label{tab:ba_tmux}
\end{table}

\subsection{Magnetic shielding}
\label{magnetic shield}

The SQUID readout system is susceptible to external magnetic fields and \biceparray's magnetic shielding architecture is designed to minimize the level of spurious signals generated as the telescope moves along the Earth's magnetic field lines.
The shielding configuration is optimized using COMSOL Multiphysics\footnote{COMSOL, Inc., Burlington, MA 01803 (\texttt{www.comsol.com})}, which allowed us to simulate the Meissner behavior of a superconducting material.

\begin{figure}
  \centering
  \includegraphics[width=.75\textwidth]{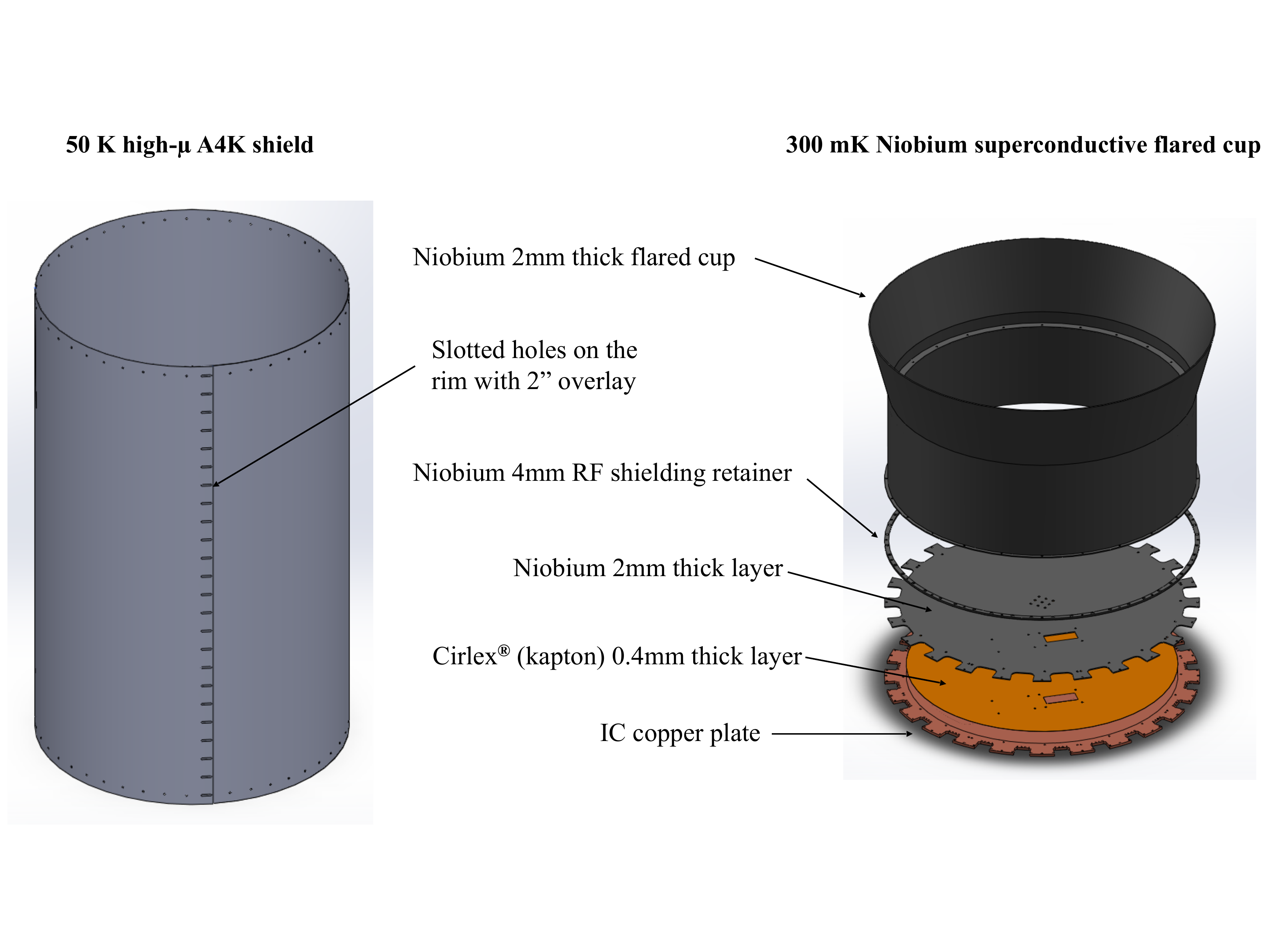}
  \caption[First stage \biceparray\ magnetic shield]{
First stage \biceparray\ magnetic shielding. A 1\,mm thick, high-$\mu$ A4K sheet is wrapped around the 50\,K stage. A multi-stage, niobium flared cup shield is mounted onto the 350\,mK stage.
}
  \label{fig:ba_mag_shield}
  \index{figure}
\end{figure}

The first stage of shielding is achieved by the combination of a 1\,mm thick high-$\mu$ A4K sheet wrapped around the 50\,K stage, and a 2\,mm thick superconducting niobium flared cup on the 350\,mK stage (Figure~\ref{fig:ba_mag_shield}).
Slotted mounting holes on the 50\,K shield allow sufficient thermal expansion mismatch between the aluminum and A4K.
The sub-Kelvin niobium shield required a continuous base in addition to the flared cylinder to enhance the shielding performance.
The niobium base is not electrically connected to the copper base plate to avoid potential superconductor-metal interactions at the contact surface.
The receiver-level shielding provides a $>200$ suppression factor from external magnetic fields.

The detectors and first stage SQUIDs are further protected by the module housing. 
The combination of the aluminum detector frame, with a niobium backshort and niobium enclosed-box form a semi-enclosed, superconducting environment for the readout electronics.
Simulation shows the biggest field leakage comes from the slots for the readout cables.
An additional niobium sheet, offsetting the cable slots is mounted on the backside of the module niobium housing to minimize the leakage.
To further increase the shielding, a 0.020" thick high-$\mu$ A4K sheet is placed in between the niobium backshort and the circuit boards, 1~mm away from the SQUID chips to draw the remaining, unwanted magnetic field away from the SQUIDs.
We estimate a final shielding factor in excess of 500 from the detector module.
Table~\ref{tab:ba_mag} shows the simulated magnetic shielding performance of the receiver and module.

\begin{table}
  \center
  \begin{tabular}{l c c} 
    \toprule 
    Parts				&	Axial Residual Flux 	&	Transverse Residual Flux	\\
    \midrule 
    50K and 4K shield			&	0.24\%			&	0.52\%		\\
    Detector module				&	0.12\%			&	0.20\%		\\    
    Final shielding				&	0.0003\%			&	0.001\%		\\
    \bottomrule
    \vspace{1mm}
  \end{tabular}
  \caption{Simulated magnetic shielding performance in \biceparray}
  \label{tab:ba_mag}
\end{table}

\section{Observation strategy and projected performance}

\begin{figure}
  \centering
  {\includegraphics[width=.78\columnwidth, clip=true]{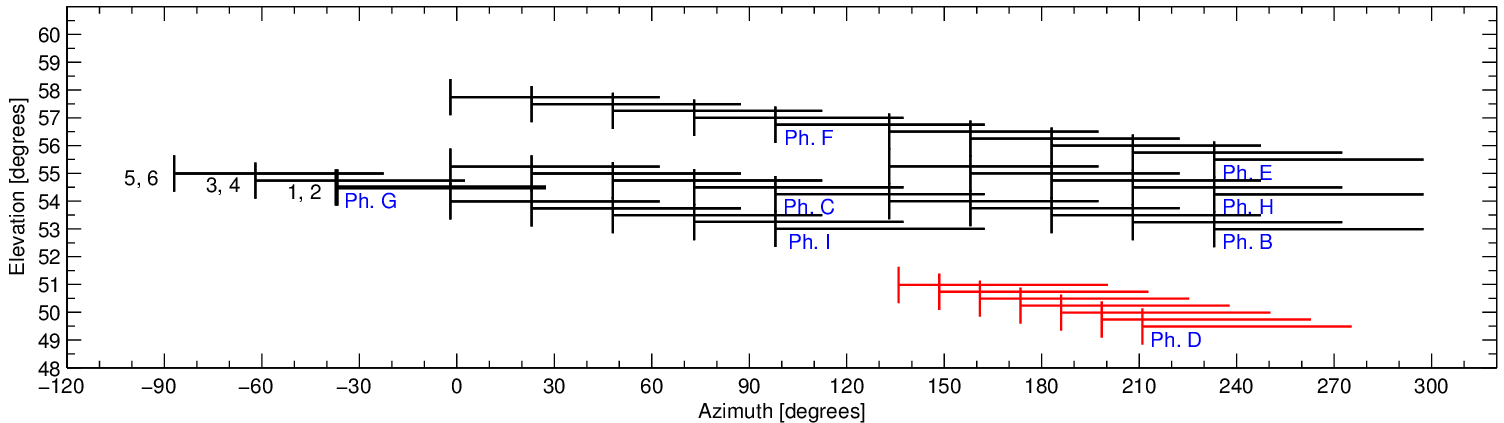}}
  {\includegraphics[width=.2\columnwidth, clip=true]{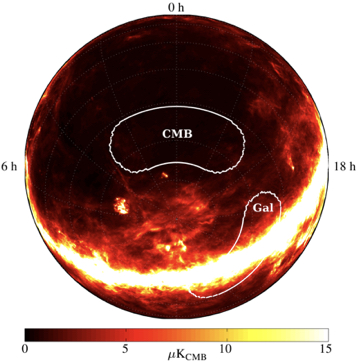}}
  \caption{{\it Left}: Observing schedule in ground-based coordinates. Horizontal lines indicate the field scans and the vertical lines indicate the bracketing elevation nods. The telescope scans at a fixed elevation during each scanset, including one galactic plane phase. {\it Right}: \bicep\ observation sky patch, with the $\sim 600\text{deg}^2$ CMB field and galactic scan (highlighted in red at the observing schedule).
}
  \label{fig:fgfig}
  \index{figure}
\end{figure}

\biceparray\ focuses its CMB observations on the same sky patch as \bicepthree, spanning right ascension $-60\deg<$RA$<60\deg$ and declination $-70\deg<\delta<-40\deg$, with an effective area of $\sim 600\text{deg}^2$ (larger than \keck's $\sim 400\text{deg}^2$ but colocated). 
With the full \biceparray\ sensitivity, we expect to achieve and surpass the \planck\ map depths at all frequencies after only a few months of observations. Figure~\ref{fig:fgfig} shows the planned observing strategy, which is identical to the current \bicepthree\ schedule.

\begin{figure}
  \centering
  \includegraphics[width=.75\textwidth]{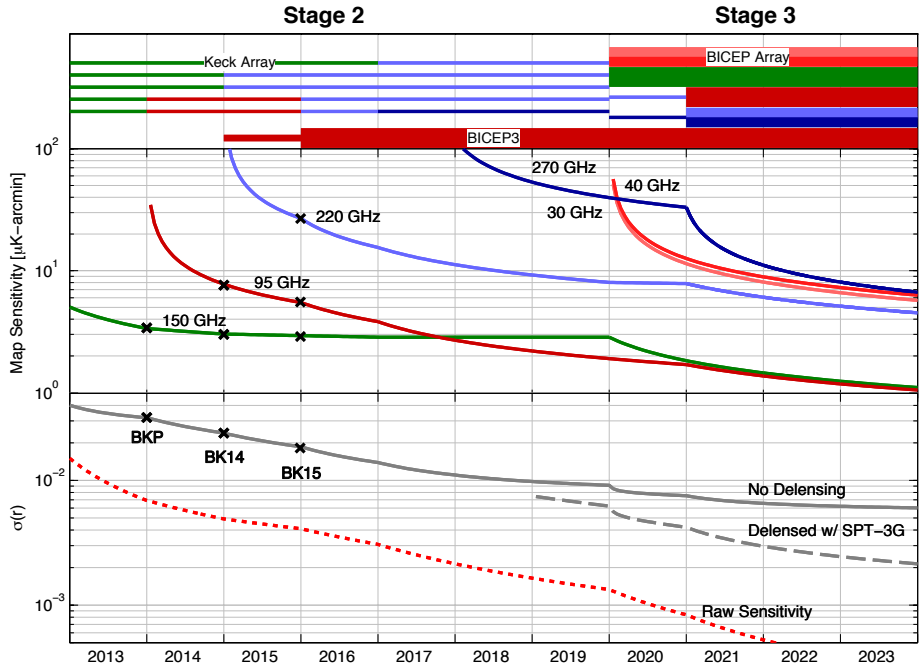}
  \caption{
  The projections on sensitivity to the tensor-to-scalar ratio that include a full estimation of lensing and Galactic foregrounds, pinned by end-to-end calculations from \bicep/\keck\ analysis for data obtained through 2015.
  }
  \label{fig:bk_projections_2018}
  \index{figure}
\end{figure}

The parameters of the \keck, \bicepthree, and \biceparray\ are given in Table~\ref{tab:rxs}.
The \biceparray\ sensitivity estimate is based on achieved survey weight per year of the \keck.
This procedure accounts for all real-world observing imperfections such as detector yield, cryogenic efficiency, data cuts, ground subtraction, temporal filtering, and other unexpected events that decrease the final sensitivity compared to the ideal case.
Figure~\ref{fig:bk_projections_2018} shows the projected sensitivity of the ongoing and planned \bicep\ program.
We expected to reach $ \sigma(r) \sim 0.003$ at the end of the program, depending on the level of delensing that can be achieved using higher-resolution data from the South Pole Telescope.

\acknowledgments 
 
The \bicep/\keck\ project have been made possible through a series of grants from the National Science Foundation including 0742818, 0742592, 1044978, 1110087, 1145172, 1145143, 1145248, 1639040, 1638957, 1638978, 1638970, \& 1726917 and by the Keck Foundation.
The development of antenna-coupled detector technology was supported by the JPL Research and Technology Development Fund and NASA Grants 06-ARPA206-0040, 10-SAT10-0017, 12-SAT12-0031, 14-SAT14-0009 \& 16-SAT16-0002.
The development and testing of focal planes were supported by the Gordon and Betty Moore Foundation at Caltech.
Readout electronics were supported by a Canada Foundation for Innovation grant to UBC.
The computations in this paper were run on the Odyssey cluster supported by the FAS Science Division Research Computing Group at Harvard University.
The analysis effort at Stanford and SLAC is partially supported by the U.S. DoE Office of Science.
We thank the staff of the U.S. Antarctic Program and in particular the South Pole Station without whose help this research would not have been possible.
Tireless administrative support was provided by Kathy Deniston, Sheri Stoll, Irene Coyle, Donna Hernandez, and Dana Volponi. 

\bibliography{report} 

\begin{thebibliography}{10}

\bibitem{kamionkowski2016quest}
Kamionkowski, M. and Kovetz, E.~D., ``The quest for b modes from inflationary
  gravitational waves,'' {\em Annual Review of Astronomy and Astrophysics}~{\bf
  54},  227--269 (2016).

\bibitem{BK15}
{BICEP2 Collaboration} and {Keck Array Collaboration}, ``{BICEP2 / Keck Array
  X: Improved Constraints on Inflationary Gravitational Waves using BICEP2/Keck
  Data up to the 2015 Season in Conjunction with WMAP/Planck},'' {\em ArXiv
  18xx.xxxxx}  (2018).

\bibitem{ade2014detection}
Ade, P.~A., Aikin, R.~W., Barkats, D., Benton, S.~J., Bischoff, C.~A., Bock,
  J.~J., Brevik, J.~A., Buder, I., Bullock, E., Dowell, C.~D., et~al.,
  ``Detection of b-mode polarization at degree angular scales by bicep2,'' {\em
  Physical Review Letters}~{\bf 112}(24),  241101 (2014).

\bibitem{BKP}
{BICEP2/Keck Collaboration}, {Planck Collaboration}, {Ade}, P.~A.~R.,
  {Aghanim}, N., {Ahmed}, Z., {Aikin}, R.~W., {Alexander}, K.~D., {Arnaud}, M.,
  {Aumont}, J., {Baccigalupi}, C., and et~al., ``{Joint Analysis of BICEP2/Keck
  Array and Planck Data},'' {\em Physical Review Letters}~{\bf 114},  101301
  (Mar. 2015).

\bibitem{kang2018bicep3}
Kang, J., Ade, P., Ahmed, Z., Alexander, K.~D., Amiri, M., Barkats, D., Benton,
  S., Bischoff, C.~A., Bock, J., Boenish, H., et~al., ``2017 upgrade and
  performance of bicep3: a 95ghz refracting telescope for degree-scale cmb
  polarization,'' in [{\em Millimeter, Submillimeter, and Far-Infrared
  Detectors and Instrumentation for Astronomy
  VIII}{\nolinebreak\hspace{0.1em}]},  International Society for Optics and
  Photonics (2018).

\bibitem{crumrine2018biceparray}
Crumrine, M., Ade, P., Ahmed, Z., Alexander, K.~D., Amiri, M., Barkats, D.,
  Benton, S., Bischoff, C.~A., Bock, J., Boenish, H., et~al., ``Bicep array
  cryostat and mount design,'' in [{\em Millimeter, Submillimeter, and
  Far-Infrared Detectors and Instrumentation for Astronomy
  VIII}{\nolinebreak\hspace{0.1em}]},  International Society for Optics and
  Photonics (2018).

\bibitem{2016JLTPB3}
{Wu}, W.~L.~K., {Ade}, P.~A.~R., {Ahmed}, Z., {Alexander}, K.~D., {Amiri}, M.,
  {Barkats}, D., {Benton}, S.~J., {Bischoff}, C.~A., {Bock}, J.~J.,
  {Bowens-Rubin}, R., {Buder}, I., {Bullock}, E., {Buza}, V., {Connors}, J.~A.,
  {Filippini}, J.~P., {Fliescher}, S., {Grayson}, J.~A., {Halpern}, M.,
  {Harrison}, S.~A., {Hilton}, G.~C., {Hristov}, V.~V., {Hui}, H., {Irwin},
  K.~D., {Kang}, J., {Karkare}, K.~S., {Karpel}, E., {Kefeli}, S.,
  {Kernasovskiy}, S.~A., {Kovac}, J.~M., {Kuo}, C.~L., {Megerian}, K.~G.,
  {Netterfield}, C.~B., {Nguyen}, H.~T., {O'Brient}, R., {Ogburn}, R.~W.,
  {Pryke}, C., {Reintsema}, C.~D., {Richter}, S., {Sorensen}, C.,
  {Staniszewski}, Z.~K., {Steinbach}, B., {Sudiwala}, R.~V., {Teply}, G.~P.,
  {Thompson}, K.~L., {Tolan}, J.~E., {Tucker}, C.~E., {Turner}, A.~D.,
  {Vieregg}, A.~G., {Weber}, A.~C., {Wiebe}, D.~V., {Willmert}, J., and {Yoon},
  K.~W., ``{Initial Performance of Bicep3: A Degree Angular Scale 95 GHz Band
  Polarimeter},'' {\em Journal of Low Temperature Physics}~{\bf 184},  765--771
  (Aug. 2016).

\bibitem{dierickx2018window}
Dierickx, M., Ade, P., Ahmed, Z., Alexander, K.~D., Amiri, M., Barkats, D.,
  Benton, S., Bischoff, C.~A., Bock, J., Boenish, H., et~al., ``Design and
  optimization of vacuum windows for the bicep/keck array cmb experiment,'' in
  [{\em Millimeter, Submillimeter, and Far-Infrared Detectors and
  Instrumentation for Astronomy VIII}{\nolinebreak\hspace{0.1em}]},
  International Society for Optics and Photonics (2018).

\bibitem{inoue2014cryogenic}
Inoue, Y., Matsumura, T., Hazumi, M., Lee, A.~T., Okamura, T., Suzuki, A.,
  Tomaru, T., and Yamaguchi, H., ``Cryogenic infrared filter made of alumina
  for use at millimeter wavelength,'' {\em Applied optics}~{\bf 53}(9),
  1727--1733 (2014).

\bibitem{Runyan2008}
{Runyan}, M.~C. and {Jones}, W.~C., ``{Thermal conductivity of
  thermally-isolating polymeric and composite structural support materials
  between 0.3 and 4 K},'' {\em Cryogenics}~{\bf 48},  448--454 (Sept. 2008).

\bibitem{ICC_Duband}
{Duband}, L., {Prouve}, T., {Bock}, J.~J., {Moncelsi}, L., and {Schillaci}, A.,
  ``Sub-kelvin cooling for the bicep array project,'' in [{\em Cryocoolers
  20}{\nolinebreak\hspace{0.1em}]},  International Cryocooler Conference, Inc.,
  Burlington, VT (2018).

\bibitem{hui2018measuring}
Hui, H., {\em Measuring the Polarization of the Cosmic Microwave Background
  with BICEP3}, PhD thesis, California Institute of Technology (2018).

\bibitem{Roger2015}
{BICEP2 Collaboration}, {Keck Array Collaboration}, {SPIDER Collaboration},
  {Ade}, P.~A.~R., {Aikin}, R.~W., {Amiri}, M., {Barkats}, D., {Benton}, S.~J.,
  {Bischoff}, C.~A., {Bock}, J.~J., {Bonetti}, J.~A., {Brevik}, J.~A., {Buder},
  I., {Bullock}, E., {Chattopadhyay}, G., {Davis}, G., {Day}, P.~K., {Dowell},
  C.~D., {Duband}, L., {Filippini}, J.~P., {Fliescher}, S., {Golwala}, S.~R.,
  {Halpern}, M., {Hasselfield}, M., {Hildebrandt}, S.~R., {Hilton}, G.~C.,
  {Hristov}, V., {Hui}, H., {Irwin}, K.~D., {Jones}, W.~C., {Karkare}, K.~S.,
  {Kaufman}, J.~P., {Keating}, B.~G., {Kefeli}, S., {Kernasovskiy}, S.~A.,
  {Kovac}, J.~M., {Kuo}, C.~L., {LeDuc}, H.~G., {Leitch}, E.~M., {Llombart},
  N., {Lueker}, M., {Mason}, P., {Megerian}, K., {Moncelsi}, L., {Netterfield},
  C.~B., {Nguyen}, H.~T., {O'Brient}, R., {Ogburn}, IV, R.~W., {Orlando}, A.,
  {Pryke}, C., {Rahlin}, A.~S., {Reintsema}, C.~D., {Richter}, S., {Runyan},
  M.~C., {Schwarz}, R., {Sheehy}, C.~D., {Staniszewski}, Z.~K., {Sudiwala},
  R.~V., {Teply}, G.~P., {Tolan}, J.~E., {Trangsrud}, A., {Tucker}, R.~S.,
  {Turner}, A.~D., {Vieregg}, A.~G., {Weber}, A., {Wiebe}, D.~V., {Wilson}, P.,
  {Wong}, C.~L., {Yoon}, K.~W., and {Zmuidzinas}, J., ``{Antenna-coupled TES
  Bolometers Used in BICEP2, Keck Array, and Spider},'' {\em Astrophys.\
  J.}~{\bf 812},  176 (Oct. 2015).

\bibitem{hui2016bicep3}
Hui, H., Ade, P., Ahmed, Z., Alexander, K.~D., Amiri, M., Barkats, D., Benton,
  S., Bischoff, C.~A., Bock, J., Boenish, H., et~al., ``Bicep3 focal plane
  design and detector performance,'' in [{\em Millimeter, Submillimeter, and
  Far-Infrared Detectors and Instrumentation for Astronomy
  VIII}{\nolinebreak\hspace{0.1em}]},   {\bf 9914},  99140T, International
  Society for Optics and Photonics (2016).

\bibitem{ahmed2018corr}
Soliman, A., Ade, P., et~al., ``Design and performance of single/wide band
  corrugated walls for the bicep array detector modules at 30/40 ghz,'' in
  [{\em Millimeter, Submillimeter, and Far-Infrared Detectors and
  Instrumentation for Astronomy VIII}{\nolinebreak\hspace{0.1em}]},
  International Society for Optics and Photonics (2018).

\bibitem{Battistelli2008}
{Battistelli}, E.~S., {Amiri}, M., {Burger}, B., {Halpern}, M., {Knotek}, S.,
  {Ellis}, M., {Gao}, X., {Kelly}, D., {Macintosh}, M., {Irwin}, K., and
  {Reintsema}, C., ``{Functional Description of Read-out Electronics for
  Time-Domain Multiplexed Bolometers for Millimeter and Sub-millimeter
  Astronomy},'' {\em Journal of Low Temperature Physics}~{\bf 151},  908--914
  (May 2008).

\bibitem{obrient2018tkid}
O'Brient, R. et~al., ``Thermal kinetic inductance detectors for cmb and sub
  millimeter observations,'' in [{\em Millimeter, Submillimeter, and
  Far-Infrared Detectors and Instrumentation for Astronomy
  VIII}{\nolinebreak\hspace{0.1em}]},  International Society for Optics and
  Photonics (2018).

\bibitem{henderson2018umux}
Henderson, S. et~al., ``Highly-multiplexed microwave squid readout using the
  slac microresonator radio frequency (smurf) electronics for future cmb and
  sub-millimeter surveys,'' in [{\em Millimeter, Submillimeter, and
  Far-Infrared Detectors and Instrumentation for Astronomy
  VIII}{\nolinebreak\hspace{0.1em}]},  International Society for Optics and
  Photonics (2018).

\end{thebibliography}
\bibliographystyle{spiebib} 

\end{document}